\documentclass[10pt,aps,pre,twocolumn,groupedaddress,longbibliography,
showpacs]{revtex4-1}
\usepackage{epsfig,amsmath,amssymb,verbatim,MnSymbol}

\begin{document}


\title{Thermodynamics of Micro- and Nano-Systems Driven by Periodic
Temperature Variations}


\author{Kay Brandner\textsuperscript{1}}
\author{Keiji Saito\textsuperscript{2}}
\author{Udo Seifert\textsuperscript{1}}
\affiliation{\textsuperscript{{{\rm 1}}}II. Institut f\"ur Theoretische Physik, Universit\"at Stuttgart, 70550 Stuttgart, Germany\\
\textsuperscript{{{\rm 2}}}Department of Physics, Keio University, 3-14-1 Hiyoshi, Kohoku-ku, Yokohama, Japan 223-8522}


\date{\today}

\begin{abstract}

We introduce a general framework for analyzing the thermodynamics of small systems that are driven by both a periodic temperature variation and some external parameter modulating their energy. This set-up covers, in particular, periodic micro and nano-heat engines. In a first step, we show how to express total entropy production by properly identified time-independent affinities and currents without making a linear response assumption. In linear response, kinetic coefficients akin to Onsager coefficients can be identified. Specializing to a Fokker-Planck type dynamics, we show that these coefficients can be expressed as a sum of an adiabatic contribution and one reminiscent of a Green-Kubo expression that contains deviations from adiabaticity. Furthermore, we show that the generalized kinetic coefficients fulfill an Onsager-Casimir type symmetry tracing back to microscopic reversibility. This symmetry allows for non-identical off-diagonal coefficients if the driving protocols are not symmetric under time-reversal. We then derive a novel constraint on the kinetic coefficients that is sharper than the second law and provides an efficiency-dependent bound on power. As one consequence, we can prove  that the power vanishes at least linearly when approaching Carnot efficiency. We illustrate our general framework by explicitly working out the paradigmatic case of a Brownian heat engine realized by a colloidal particle in a time-dependent harmonic trap subject to a periodic temperature profile. This case study reveals inter alia that our new general bound on power is asymptotically tight.

\end{abstract}


\maketitle

\newcommand{\figref}[1]{Fig. \ref{#1}}
\newcommand{\secref}[1]{Sec. #1}
\newcommand{\F}{\mathcal{F}}
\newcommand{\B}{\mathbf{B}}
\newcommand{\x}{\mathbf{x}}
\newcommand{\T}{\mathcal{T}}
\renewcommand{\L}{\mathsf{L}}
\newcommand{\J}{\mathcal{J}}
\newcommand{\be}{\boldsymbol{\varepsilon}}
\newcommand{\e}{\varepsilon}
\newcommand{\plc}{p^{{\rm c}}(\x,t)}
\newcommand{\xint}{\int \!\! d^n \x \;}
\renewcommand{\tint}{\int_0^{\T}\!\!\!\! dt \;}
\newcommand{\tauint}{\int_0^{\infty} \!\!\!\! d\tau \;}
\newcommand{\taupint}{\int_0^{\infty}\!\!\!\! d\tau'\;}
\newcommand{\ttauint}{
\int_0^\T\!\!\!\! dt \int_0^\infty\!\!\!\! d\tau \;}
\newcommand{\txint}{\tint\!\!\!\xint}
\newcommand{\peq}{\bar{p}(\x,t)}
\newcommand{\pq}{p^{{\rm eq}}(\x)}
\newcommand{\ev}[1]{\left\langle #1 \right\rangle}
\newcommand{\txtauint}{\tint\!\!\!\xint\!\!\!\tauint}
\newcommand{\tL}{\tilde{\L}}
\newcommand{\D}{\Delta}
\newcommand{\ea}{\emph{et al.~}}
\newcommand{\bt}{\bar{\eta}}
\renewcommand{\k}{\kappa}
\newcommand{\K}{\L^{0\dagger}}
\newcommand{\tK}{\tilde{\L}^{0\dagger}}
\newcommand{\y}{\mathbf{y}}
\newcommand{\g}{\gamma}
\newcommand{\gt}{\gamma}
\renewcommand{\l}{\lambda}
\newcommand{\ta}{\gamma_q}
\newcommand{\ps}[1]{\left\llangle #1 \right\rrangle}
\renewcommand{\d}{\delta}
\newcommand{\dg}{\d\dot{g}}

\vbadness=10000
\hbadness=10000

\section{Introduction}

Thermodynamic processes on the micro- and nano-scale in systems driven
out of equilibrium by periodically changing control parameters like an
external force or the temperature of their environment, can
be scrutinized under the microscope by virtue of precise measurements
of characteristic quantities such as applied work or exchanged heat
\cite{Steeneken2010,Blickle2011,Ribezzi-Crivellari2014,Koski2014,
Pekola2015,Martinez2015}.
Despite such groundbreaking results, so far, no general theoretical
framework for the thermodynamic description of periodically driven
systems beyond the quasi-static limit is available.

In principle, the concepts of irreversible thermodynamics, a
phenomenological but powerful theory, which, building on the principle
of local equilibrium, furnishes non-equilibrium steady states with a
universal thermodynamics structure \cite{Callen1985}, can be
transferred to periodic systems \cite{Izumida2009,Izumida2010,
Izumida2015}. 
These results are, however, crucially tied to specific models and
require rather involved and nonintuitive definitions for currents
and affinities. 

In this paper, we overcome these limitations.
Starting from the first law formulated for an arbitrary system driven
out of equilibrium by both, periodic perturbations of its Hamiltonian
and the temperature of a surrounding heat bath, we obtain natural and
perfectly general identifications of fluxes and affinities.
Since they are defined on the level of cycle averages, these 
quantities are time-independent. 
Nevertheless, our new formalism captures essential finite-time 
properties of the driven system and permits a discussion of quantities
like power, which are out of reach for the laws of classical 
thermodynamics due to its notorious lack of time scales. 

Moreover, our approach provides a universal prescription for the 
construction of kinetic coefficients, which fully characterize the
system in the linear response regime. 
By using a well established and rather general stochastic approach
based on a Fokker-Planck equation to describe the underlying 
time-dependent dynamics \cite{Jung1993}, we prove that these
coefficients are interrelated by a remarkable symmetry, which, just
like Onsager's celebrated reciprocity relations \cite{Onsager1931,
Onsager1931a}, can be traced back to microscopic reversibility. 
 
Ever since James Watt's steam engine, the urge to explore the
fundamental principles governing the performance of cyclic heat
engines, was one of the major quests in thermodynamics. 
Two key figures of merit are particularly important in this context.
While efficiency, the first one, is universally bounded by the 
Carnot value as a direct consequence of the second law, similar
constraints on power, the second one, could, so far, be obtained 
only within specific setups, see for example \cite{Schmiedl2008,
Esposito2010a,Abe2011,Allahverdyan2013,Holubec2014}.

Thermoelectric heat engines, which, in contrast to cyclic ones, work
in a steady state \cite{Humphrey2005a,Benenti2013} are likewise 
subject of substantial research efforts concerning efficiency 
\cite{Humphrey2002,Humphrey2005}, power \cite{Whitney2014,
Whitney2014a} and the relation between these quantities
\cite{Esposito2009a,Sothmann2015}.
At least in the linear response regime, however, much more general
results are currently available for this type of heat engines than for
their periodic counterparts.
Specifically, it can be shown that, in the linear regime, the power of
such devices is bounded by a quadratic function of efficiency that
vanishes at the Carnot value $\eta_{{\rm C}}$ and attains its maximum
at the Curzon-Ahlborn value $\eta_{{\rm C}}/2$ \cite{Curzon1975,
VandenBroeck2005,Esposito2009,Seifert2012}.
This result follows from a quite general analysis within the
framework of linear irreversible thermodynamics. 
It must, however, be reconsidered in the presence of a magnetic field,
which breaks the Onsager symmetry, an issue which is currently under
active discussion \cite{Benenti2011,Saito2011,Sanchez2011,
Horvat2012,Benenti2013,Brandner2013,Brandner2013a,Balachandran2013,
Stark,Brandner2015,Sabass2015}. 

Here, by applying our new formalism, we establish the
aforementioned quadratic bound on power for cyclic heat engines in
linear response. 
In particular, by using a novel method, which does not require any
additional assumptions, we prove that this bound still holds if the 
matrix of kinetic coefficients is not symmetric, which is the generic
case for periodically driven systems. 
We emphasize that, by now, an analogous result for thermoelectric 
heat engines is not available on such a general level.
To complete our analysis, we show that this bound on power is tight
within a paradigmatic model of a Brownian heat engine, which was
originally proposed in \cite{Schmiedl2008} and recently realized in a
landmark experiment \cite{Blickle2011}. 

The rest of this paper is structured in three major parts. 
In section II, we develop our general formalism and prove a
generalized reciprocity relation for the kinetic coefficients of 
periodic systems.
Section III is devoted to the discussion of cyclic heat engines in 
the linear response regime and the derivation of a general bound on 
power. We illustrate our results by considering a simple model 
system in section IV.
Finally, we conclude in section V. 

\section{Framework}\label{Sec_Framework}
In this section, we will demonstrate that the notions of 
irreversible thermodynamics can be transferred to periodically 
driven systems. 

\subsection{Nonlinear Regime}
\begin{figure}
\epsfig{file=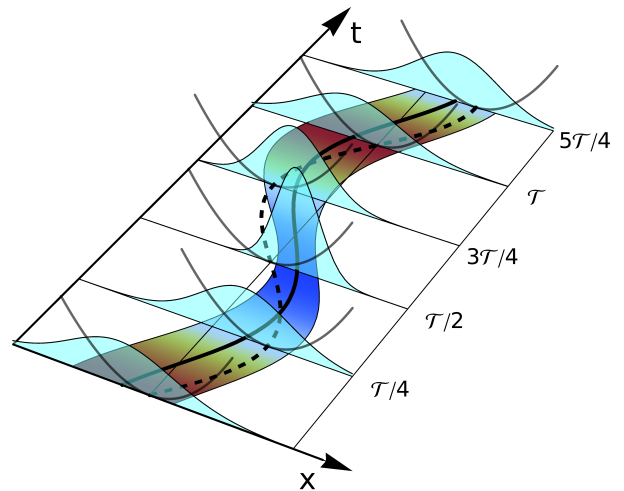,scale=0.066}
\caption{Non-equilibrium periodic state of a system with one degree of
freedom $x$ in a sinusoidally shifted harmonic potential, represented 
by gray parabolas, which is embedded in an environment with 
periodically changing temperature as indicated by the periodic color
gradient.
The solid line in the $x$-$t$ plane shows the motion of the center of
the Gaussian phase space distribution, which lags behind position of
the potential minimum shown as dashed line.  
At any time $t$, the width of the colored region equals the width of
the phase space distribution, which varies according to the
temperature.
\label{Fig_Fancy}}
\end{figure}
We begin with a brief review of the energetics of driven systems
in thermal contact with a heat bath \cite{Seifert2012}. 
Specifically, we consider a classical system with degrees of 
freedom $\x\equiv (x_1,\dots,x_n)$ and time-dependent Hamiltonian 
\begin{equation}\label{SHE_HDecomp}
H(\x,t) \equiv H_0(\x) + \Delta H g_w(\x,t),
\end{equation}
which is immersed in a heat bath, whose temperature $T(t)$ oscillates
between the two values $T_{{\rm c}}$ and $T_{{\rm h}}>T_{{\rm c}}$.
Here, $g_w(\x,t)$ denotes an externally controlled dimensionless 
function of order $1$ and $\Delta H$ quantifies the strength of this
time-dependent perturbation. 
The power extracted by the controller thus reads 
\begin{equation}\label{SHE_InstWork}
\dot{W}(t)\equiv - \xint \dot{H}(\x,t)p(\x,t),
\end{equation}
where $p(\x,t)$ denotes the  normalized probability density to find 
the system in the state $\x$ at the time $t$ and dots indicate time 
derivatives throughout the paper.
To compensate for this loss in internal energy
\begin{equation}\label{SHE_InstIntEnergy}
U(t) = \xint H(\x,t)p(\x,t),
\end{equation}
the system picks up the heat 
\begin{equation}\label{SHE_InstHeat}
\dot{Q}(t)\equiv \xint H(\x,t)\dot{p}(\x,t). 
\end{equation}
from the environment as stipulated by the first law
\begin{equation}\label{SHE_FirstLaw}
\dot{U}(t)\equiv  \dot{Q}(t)-\dot{W}(t).
\end{equation}

We will now pass from time-dependent to constant variables by
exploiting the periodic boundary conditions
\begin{equation}\label{SHE_PerBoundCond}
H(\x,t+\T) = H(\x,t) \quad\text{and}\quad
T(t+\T) = T(t),
\end{equation}
where $\T$ is the length of one operation cycle. 
Quite naturally,  we invoke the assumption that, given these
conditions, the time evolution of the probability density $p(\x,t)$
eventually converges to a periodic limit
\begin{equation}\label{SHE_PerLimitDist}
\plc = p^{{\rm c}}(\x,t+\T)
\end{equation} 
as illustrated in \figref{Fig_Fancy} for a simple system.
Once this periodic state is reached, the average entropy production
per cycle arises solely due to heat exchange between the system and
the environment, since, due to the periodicity of the the distribution
$\plc$, no net entropy is generated in the system in a full cycle, 
i.e., we have
\begin{equation}\label{SHE_EntProdDef}
\dot{S} =  -\frac{1}{\T}\tint \frac{\dot{Q}(t)}{T(t)}.
\end{equation}
By inserting \eqref{SHE_InstHeat} into \eqref{SHE_EntProdDef} and 
parameterizing $T(t)$ by a dimensionless function $0\leq \ta(t)\leq 1$
such that
\begin{equation}
T(t)\equiv \frac{T_{{\rm c}}T_{{\rm h}}}
{T_{{\rm h}}+ (T_{{\rm c}}-T_{{\rm h}})\ta(t)},
\end{equation}
it is straightforward to derive the expression 
\begin{multline}\label{SHE_EntProdIntm}
\dot{S}  =  \frac{\Delta H}{T_{{\rm c}}}
            \frac{1}{\T}\txint\dot{g}_w(\x,t)\plc\\
            +\left(\frac{1}{T_{{\rm c}}}-\frac{1}{T_{{\rm h}}}\right)
            \frac{1}{\T}\txint \ta(t)H(\x,t)\dot{p}^{{\rm c}}(\x,t)
\end{multline}
using one integration by parts with respect to time. 
The corresponding boundary terms vanish due to the periodicity of 
the involved quantities.
Obviously \eqref{SHE_EntProdIntm} can be cast
in the generic form \cite{Callen1985}
\begin{equation}\label{SHE_EntProdGen}
\dot{S} = \F_w J_w + \F_q J_q,
\end{equation}
by identifying the work flux
\begin{equation}\label{SHE_WorkFluxDef}
J_w\equiv\frac{1}{\T}\txint\dot{g}_w(\x,t)\plc,
\end{equation}
the generalized heat flux
\begin{equation}\label{SHE_HeatFluxDef}
J_q\equiv \frac{1}{\T}\txint \ta(t)H(\x,t)\dot{p}^{{\rm c}}(\x,t)
\end{equation}
and the affinities 
\begin{equation}
\F_w\equiv \Delta H/T_{{\rm c}}
\quad\text{and}\quad
\F_q\equiv 1/T_{{\rm c}}-1/T_{{\rm h}}.
\end{equation}
Within only a few lines, we have thus obtained our first main result,
namely, we recovered for periodically time-dependent systems the
structure of irreversible thermodynamics.
The key point here is the identification of appropriate pairs of 
affinities and fluxes, whose respective products sum up to the total
entropy production. 

For later purposes, we note that, after one integration by part with
respect to $t$, the heat flux \eqref{SHE_HeatFluxDef} can be rewritten
as 
\begin{multline}\label{SHE_HeatFluxMeanRep}
J_q =\frac{1}{\T} \txint \dot{g}_q \plc\\
    +\frac{\Delta H}{\T}\txint\g_q(t)g_w(\x,t)\dot{p}^{{\rm c}}(\x,t)
\end{multline}
where 
\begin{equation}\label{SHE_gqDef}
g_q(\x,t)\equiv - H_0(\x)\ta(t).
\end{equation}

\subsection{Linear Response Regime}\label{Sec_OC}

\subsubsection{Kinetic Coefficients}
We now focus on the linear regime with respect to the temporal 
gradients $\Delta H$ and $\Delta T\equiv T_{{\rm h}}-T_{{\rm c}}$.
By expanding the fluxes \eqref{SHE_WorkFluxDef} and 
\eqref{SHE_HeatFluxDef}, we obtain
\begin{equation}\label{LSH_PhenEq}
\begin{split}
J_w &= L_{ww}\F_w + L_{wq}\F_q + \mathcal{O}(\Delta^2),\\
J_q &= L_{qw}\F_w + L_{qq}\F_q + \mathcal{O}(\Delta^2)
\end{split}
\end{equation}
with linearized affinities
\begin{equation}
\F_w = \Delta H/T_{{\rm c}} \quad\text{and}\quad
\F_q = \Delta T/T_{{\rm c}}^2 + \mathcal{O}\left(\Delta^2\right)
\end{equation}
and kinetic coefficients 
\begin{equation}\label{LSH_OCDef}
L_{\alpha\beta}\equiv 
\left.\frac{\partial J_\alpha}{\partial\F_{\beta}}\right|_{\F=0} 
\quad\text{for}\quad
\alpha,\beta=w,q.
\end{equation}
The entropy production \eqref{SHE_EntProdDef} thus reduces to 
\begin{equation}\label{LSH_EntProd}
\dot{S}=\sum_{\alpha,\beta=w,g} L_{\alpha\beta}\F_\alpha\F_\beta.
\end{equation}
To guarantee that this expression is nonnegative for any $\F_\alpha$
as stipulated by the second law, the kinetic coefficients must obey 
the constraints 
\begin{equation}\label{LSH_SecondLaw}
L_{ww}, L_{qq}\geq 0 \;\;\; \text{and}\;\;\;
L_{ww}L_{qq} - (L_{wq}+L_{qw})^2/4\geq 0,
\end{equation}
which we prove explicitly in \secref{\ref{Sec_SHEBoundsPowEff}}
for a large class of systems.
It is, however, not evident at this stage whether a reciprocity 
relation relating $L_{\alpha\beta}$ with $L_{\beta\alpha}$ or any
further constraints exist. 

\subsubsection{Adiabatic Limit}

As a first step, we investigate the adiabatic regime, which is
characterized by the Hamiltonian $H(\x,t)$ and the temperature $T(t)$
changing slowly enough in time such that the system effectively
passes through a sequence of equilibrium states, i.e., 
\begin{equation}\label{LSH_AdDist}
\plc=\exp\left[-H(\x,t)/(k_BT(t))\right]/Z(t) 
\end{equation}
with
\begin{equation}
Z(t)\equiv \xint \exp\left[-H(\x,t)/(k_BT(t))\right]
\end{equation}
and $k_B$ denoting Boltmann's constant. 
Expanding \eqref{LSH_AdDist} to linear order in $\Delta H$ and 
$\Delta T$ and inserting the result into \eqref{SHE_WorkFluxDef},
\eqref{SHE_HeatFluxMeanRep} and \eqref{LSH_OCDef} gives the universal
expression 
\begin{equation}\label{LSH_AdOC}
L_{\alpha\beta}^{{\rm ad}}= -\frac{1}{k_B}
\ps{\d\dot{g}_\alpha \d g_\beta}
\end{equation} 
for the adiabatic kinetic coefficients.
Here, we introduced the notations 
\begin{equation}\label{LSH_AvDef}
\ps{A}\equiv \frac{1}{\T}\tint \ev{A(t)}
      \equiv \frac{1}{\T}\txint A(\x,t)\pq
\end{equation}
and 
\begin{equation}\label{LSH_FluctDef}
\d A(\x,t)\equiv A(\x,t) -\ev{A(\x,t)}
\end{equation}
for any quantity $A(\x,t)$ and the equilibrium distribution of the 
unperturbed system
\begin{equation}\label{LSH_EqDist}
\pq\equiv\exp\left[-H_0(\x)/(k_BT_{{\rm c}})\right]/Z_0
\end{equation}
with $Z_0$ denoting the canonical partition function.

Notably, the coefficients \eqref{LSH_AdOC} are fully antisymmetric, 
i.e., 
\begin{equation}\label{LSH_AdContbAntiSym}
L^{{\rm ad}}_{\alpha\beta}=-L^{{\rm ad}}_{\beta\alpha}.
\end{equation}
As might be expected, this property, which can be proven by a simple 
integration by parts with respect to $t$, implies vanishing entropy 
production \eqref{LSH_EntProd} in the adiabatic limit.
This avoidance of dissipation can, however, be only achieved for an
infinite cycle duration $\T$ and therefore inevitably
comes with vanishing fluxes $J_\alpha$.

\subsubsection{Stochastic Dynamics}

For further investigations of the kinetic coefficients, we have to 
specify the dynamics, which governs the time evolution of the 
probability density $p(\x,t)$.  
Having in mind, in particular, mesoscopic systems surrounded by a 
fluctuating medium, a suitable choice is given by the Fokker-Planck 
equation
\cite{Risken1996}
\begin{equation}\label{LSH_FPEq}
\partial_t p(\x,t) = \L(\x,t)p(\x,t)
\end{equation}
with
\begin{equation}\label{LSH_FPOperator}
\L(\x,t)\equiv - \partial_{x_i} D_i(\x,H,T)
               + \partial_{x_i}\partial_{x_j}D_{ij}(\x,H,T),
\end{equation}
where summation over identical indices is understood and natural 
boundary conditions are assumed.
The Hamiltonian $H(\x,t)$ and the temperature $T(t)$ enter via the
drift and diffusion coefficients $D_i(\x,H,T)$ and $D_{ij}(\x,H,T)$, 
which thus become implicitly time-dependent 
\footnote{It is well known that \eqref{LSH_FPEq} leads to a unique, 
periodic distribution $\plc$ in the long time limit if the diffusion
matrix is strictly positive definite \cite{Owedyk1985,Jung1993}. 
However, it is readily seen that this assertion still holds for
physically meaningful scenarios with singular diffusion matrix
such as the underdamped dynamics described by Kramer's equation
\cite{Risken1996}.}.

We will now formulate a set of conditions on the general
Fokker-Planck operator \eqref{LSH_FPOperator} to adapt it to the 
physical situation that we wish to discuss here.
Since micro-reversibility plays a crucial role in linear irreversible
thermodynamics, we have to ensure that our theory complies with this 
fundamental principle.
To this end, first, at any time $t$ any possible state $\x$ of the 
system must be associated with the same energy as the time-reversed
state $\be\x\equiv(\e_1 x_1,\dots,\e_n x_n)$ with $\e_i=1$ for even 
and $\e_i=-1$ for odd variables, i.e., 
\begin{equation}\label{LSH_HamiltonianTRS}
H(\x,t)= H(\be\x,t),
\end{equation}
where, throughout the paper, the transformation $\x\rightarrow\be\x$ 
is meant to include the reversal of external magnetic fields.
Second, the unperturbed Fokker-Planck operator 
$\L^0(\x)\equiv\left. \L(\x,t)\right|_{\Delta =0}$ must obey the 
detailed balance condition \cite{Risken1996}
\begin{equation}\label{LSH_DB}
\L^0(\x)\pq = \pq \L^{0\dagger}(\be\x), 
\end{equation}
for the canonical distribution \eqref{LSH_EqDist}, which uniquely
satisfies
\begin{equation}\label{LSH_UnpertStatDist}
\L^0(\x)\pq = 0.
\end{equation}
The dagger showing up in \eqref{LSH_DB}, from here onwards,
designates the adjoint of the respective operator.
Note that, while in \eqref{LSH_UnpertStatDist} the operator $\L^0(\x)$
acts on the function $\pq$, \eqref{LSH_DB} is to be read as an 
operator identity becoming meaningful when applied to a specific
function of $\x$.
Physically, the relation \eqref{LSH_DB} means that, once the system 
has reached its equilibrium state, the rate of transitions from the
state $\x$ to the state $\x'$ is balanced by the rate of transitions
in the reverse direction.

The equilibrium Fokker-Planck operator $\L^0(\x)$ can be naturally
decomposed in a reversible and an irreversible contribution
\begin{align}
& \L_{{\rm rev}}^0(\x)  \equiv (\L^0(\x)-\L^0(\be\x))/2,\\
& \L_{{\rm irr}}^0(\x)  \equiv (\L^0(\x)+\L^0(\be\x))/2,
\end{align}
which are characterized by their respective behavior under time 
reversal. 
While the irreversible part accounts for dissipative effects induced 
by the presence of the heat bath, the reversible part describes the 
intrinsic coupling of the system's degrees of freedom, which is not 
directly affected by the fluctuating environment.
Since this autonomous part of the dynamics should preserve the 
internal energy of the system, we have to impose the condition 
\begin{equation}\label{LSH_AutDyn}
\L_{{\rm rev}}^{0\dagger}(\x)H_0(\x) = 0.
\end{equation}
We note that this consideration does not play a role in the overdamped
limit, within which the entire time evolution of the system is 
effectively irreversible due to strongly dominating friction forces.

The notion of detailed balance can not be immediately generalized
to situations with external driving and time-dependent temperature. 
However, in analogy with \eqref{LSH_UnpertStatDist}, the full
Fokker-Planck operator $\L(\x,t)$ can still be characterized by the
weaker property
\begin{equation}\label{LSH_InstEq}
\L(\x,t)\exp\left[-H(\x,t)/(k_BT(t))\right] = 0,
\end{equation}
which is naturally obeyed in the absence of nonconservative forces and
guarantees that the system follows the correct thermal equilibrium
state if the Hamiltonian and the temperature are varied 
quasi-statically.

\subsubsection{Finite-Time Kinetic Coefficients}
In the linear response regime, the Fokker-Planck operator $\L(t)$ 
showing up in \eqref{LSH_FPEq} can be replaced by the expansion
\begin{equation}\label{LSH_ExpFPO}
\L(t) \equiv \L^0 + \Delta H\L^H(t) + \Delta T \L^T(t) +\mathcal{O}(\Delta^2),
\end{equation}
where, for simplicity, from \eqref{LSH_ExpFPO} onwards, we
notationally suppress the dependence of any operator on $\x$, whenever
there is no need to indicate it explicitly.
The Fokker-Planck equation \eqref{LSH_FPEq} can then be solved with
due consideration of the boundary condition \eqref{SHE_PerBoundCond}
by treating $\L^H(t)$ and $\L^T(t)$ as first order perturbations. 
The result of this standard procedure \cite{Risken1996,Kubo1998} reads 
\begin{multline}\label{LSH_LinSol}
\plc =\pq\\
+ \sum_{X=H,T} \Delta X \! \tauint e^{\L^0\tau}\L^X(t-\tau)\pq
+ \mathcal{O}(\Delta^2).
\end{multline}
After some algebra involving condition \eqref{LSH_InstEq}, which we 
relegate to appendix \ref{Apx_DevOC} for convenience, this solution 
leads to the compact expression
\begin{equation}\label{LSH_OCJJCorr}
L_{\alpha\beta} =L^{{\rm ad}}_{\alpha\beta}
                 +\frac{1}{k_B}\tauint\ps{\d\dot{g}_\alpha(0);
                 \d\dot{g}_\beta(-\tau)}
\end{equation}
for the kinetic coefficients, where the generalized equilibrium
correlation function is defined as 
\begin{widetext}
\begin{equation}\label{LSH_EqCorr}
\ps{A(t_1);B(t_2)}\equiv\frac{1}{\T}\txint
\begin{cases}
 A(\x,t_1+t)e^{\L^0(t_1-t_2)}B(\x,t_2+t)\pq
 &\text{for}\quad(t_1\geq t_2)\\
 B(\x,t_2+t)e^{\L^0(t_2-t_1)}A(\x,t_1+t)\pq
 &\text{for}\quad(t_1<t_2)
\end{cases}
\end{equation}
\end{widetext}
for arbitrary quantities $A(\x,t)$ and $B(\x,t)$.
We recall the definition \eqref{LSH_FluctDef} of the 
$\delta$-notation.

The expression \eqref{LSH_OCJJCorr} admits an illuminating physical
interpretation. 
It shows that the kinetic coefficients of periodically driven can be
decomposed into an adiabatic contribution independently identified in
\eqref{LSH_AdOC} and a finite-time correction, which has the form of 
an equilibrium correlation function.
This result might therefore be regarded as a generalization of the 
well established Green-Kubo relations, which relate linear transport 
coefficients like electric or thermal conductivity to equilibrium
correlation functions of the corresponding currents \cite{Kubo1998}.
In our case, the role of the currents is played by the fluctuation
variables $\d\dot{g}_\alpha(\x,t)$ and the ensemble average is 
augmented by a temporal average over one operation cycle. 
We note that a similar expression has been obtained for the special 
case of the effective diffusion constant of a periodically rocked 
Brownian motor in \cite{Machura2005}. 

\subsubsection{Reciprocity Relations}

Time-reversal symmetry of microscopic dynamics appears as the 
detailed balance condition on the level of the Fokker-Planck 
equation \eqref{LSH_DB}. 
By using this relation and the Green-Kubo type formula
\eqref{LSH_OCJJCorr}, it is straightforward to derive the 
generalized reciprocity relation 
\begin{equation}\label{LSH_RcpRel}
L_{\alpha\beta}\left[H(\x,t),T(t),\B\right] 
=L_{\beta\alpha}\left[H(\x,-t),T(-t),-\B\right],
\end{equation}
where, the Onsager coefficients are considered as functionals of the
time-dependent Hamiltonian and temperature and an external magnetic
field $\B$. 
The technical details of the derivation leading to the relation
\eqref{LSH_RcpRel}, which constitutes our second main result, can be
found in appendix \ref{Apx_RcpRel}.
Here, we emphasize that, although the setup of this paper differs 
significantly from the one Onsager dealt with in his pioneering work
\cite{Onsager1931a, Onsager1931}, the symmetry \eqref{LSH_RcpRel}, 
which constitutes our second main result, and the original Onsager 
relations share microscopic reversibility as the common physical
origin. 
Since, in the presence of time-dependent driving, full time reversal
especially includes reversal of the driving protocols, naturally,
these reversed protocols show up in \eqref{LSH_RcpRel}. 

The symmetry \eqref{LSH_RcpRel} holds individually for both, the
adiabatic kinetic coefficients and the finite-time correction showing
up in \eqref{LSH_OCJJCorr}.
Given the general relation \eqref{LSH_AdContbAntiSym}, it follows that
the $\L^{{{\rm ad}}}_{\alpha\beta}$ must vanish if the driving 
protocols are symmetric under time-reversal. 

The additional relation 
\begin{equation}\label{LSH_SpRcpRel}
L_{\alpha\beta}\left[\gt_w(t),\gt_q(t),\B\right] = 
L_{\beta\alpha}\left[\gt_q(t),\gt_w(t),-\B\right],
\end{equation}
can be proven if the driving $g_w(\x,t)$ introduced in
\eqref{SHE_HDecomp} factorizes according to 
\begin{equation}\label{LSH_gFctzn}
g_w(\x,t) = g_w(\x)\gt_w(t).
\end{equation}
Hence, the off-diagonal kinetic coefficients change place if the
magnetic field is reversed and the respective protocols determining
the time dependence of the Hamiltonian and the temperature are
interchanged. 
This symmetry does not involve the reversed protocols.
It is, however, less universal than \eqref{LSH_RcpRel}, since it
requires the special structure \eqref{LSH_gFctzn}, see appendix 
\ref{Apx_RcpRel} for details. 

\section{Cyclic Stochastic Heat Engines in Linear Response}

As a key application of our new approach we will discuss the 
performance of stochastic heat engines.

\subsection{Power and Efficiency}
The main two benchmark parameters here are, first, the average power
output per operation cycle
\begin{equation}\label{BPE_Power}
P\equiv -\frac{1}{\T}\txint \dot{H}(\x,t)\plc =
-T_{{\rm c}}\F_w J_w
\end{equation}
and, second, the efficiency 
\begin{equation}\label{BPE_Efficiency}
\eta\equiv P/J_q = - T_{{\rm c}}\F_w J_w/J_q,
\end{equation}
which is bounded by the Carnot value $\eta_{{\rm C}} = 1-T_{{\rm
c}}/T_{{\rm h}}$ as a direct consequence of the second law $\dot{S}
\geq 0$.
The latter figure, which is naturally suggested by the representation
\eqref{SHE_EntProdGen} of the entropy production per cycle, should be
regarded as a generalization of the conventional thermodynamic
efficiency defined for a heat engine operating between two reservoirs
of respectively constant temperature.
Our formalism includes this scenario as the special case where
$\gt_q(t)$ is chosen as a step function
\begin{equation}\label{LSH_StepProt}
\gt_q (t) = \begin{cases}
1 &\quad\text{for}\quad 0\leq t < \T_1\\
0 &\quad\text{for}\quad \T_1\leq t < \T
\end{cases}
\end{equation}
with $0<\T_1<\T$ such that the system is in contact with the hot
temperature $T_{{\rm h}}$ in the first part of the cycle and the cold
temperature $T_{{\rm c}}$ in the second one.

Under linear response conditions, which we will assume for the rest of
this section, the fluxes $J_\alpha$ can be eliminated using 
\eqref{LSH_PhenEq} such that the expressions \eqref{BPE_Power} and
\eqref{BPE_Efficiency} reduce to 
\begin{equation}\label{BPE_PowerLin}
P = - T_{{\rm c}}\F_w\left(L_{ww}\F_w + L_{wq}\F_q\right)
\end{equation}
and 
\begin{equation}\label{BPE_EfficiencyLin}
\eta = 
-\frac{T_{{\rm c}}\F_w(L_{ww}\F_w+L_{wq}\F_q)}{L_{qw}\F_w+L_{qq}\F_q},
\end{equation}
respectively. 
Clearly, these figures are crucially determined by the kinetic
coefficients $L_{\alpha\beta}$.
In contrast to the thermoelectric case, where the reciprocity relation
$L_{\alpha\beta}=L_{\beta\alpha}$ holds without magnetic fields,
the analysis of the preceding section has revealed that
for cyclic heat engines this symmetry is typically broken if the 
driving protocols are not invariant under time reversal. 
As pointed out by Benenti \ea  \cite{Benenti2011}, a non symmetric 
matrix of kinetic coefficients leads to profound consequences for the
performance of thermoelectric devices including the option of Carnot
efficiency at finite power. 

These results apply similarly to the systems considered here, since 
our theoretical framework is structurally equivalent to the standard
theory of linear irreversible thermodynamics used as a starting point
in \cite{Benenti2011}.
To demonstrate this correspondence explicitly, following \cite{Benenti2011}, we define the dimensionless parameters
\begin{equation}\label{BPE_DimLParamters}
x\equiv \frac{L_{wq}}{L_{qw}} \quad\text{and}\quad
y\equiv \frac{L_{wq}L_{qw}}{L_{ww}L_{qq}-L_{wq}L_{qw}},
\end{equation}
which, due to the second law \eqref{LSH_SecondLaw}, are related by the
inequalities
\begin{equation}
h(x)\leq y\leq 0 \;\;\text{for} \;\; x<0, \;\;\;
0\leq y\leq h(x) \;\;\text{for} \;\; x\geq 0\!
\end{equation}
with 
\begin{equation}
h(x)\equiv\frac{4x}{(x-1)^2}.
\end{equation}
By optimizing \eqref{BPE_PowerLin} and \eqref{BPE_EfficiencyLin}, 
respectively, as functions of $\F_w$ the expressions
\begin{equation}
\eta_{{\rm max}}(x,y) \equiv \eta_{{\rm C}}x
\frac{\sqrt{y+1}-1}{\sqrt{y+1}+1},
\end{equation}
for maximum efficiency and
\begin{equation}
\eta^\ast (x,y)=\eta_{{\rm C}}\frac{xy}{4+2y},
\end{equation}
for efficiency at maximum power \cite{Seifert2012} are obtained, where
$\eta_{{\rm C}}\approx\D T/T_{{\rm c}}=T_{{\rm c}}\F_q$ denotes 
the Carnot efficiency in the linear regime. 
For $y= h(x)$ and $|x|\geq 0$, the maximum efficiency equals 
$\eta_{{\rm C}}$ and the efficiency at maximum power can exceed the 
Curzon-Ahlborn value $\eta_{{\rm C}}/2$
\cite{Curzon1975,VandenBroeck2005,Esposito2009,Seifert2011,
VandenBroeck2012,Seifert2012}, reaching even $\eta_{{\rm C}}$ in the 
limit $x\rightarrow\pm\infty$.
\begin{figure}
\flushleft
\epsfig{file=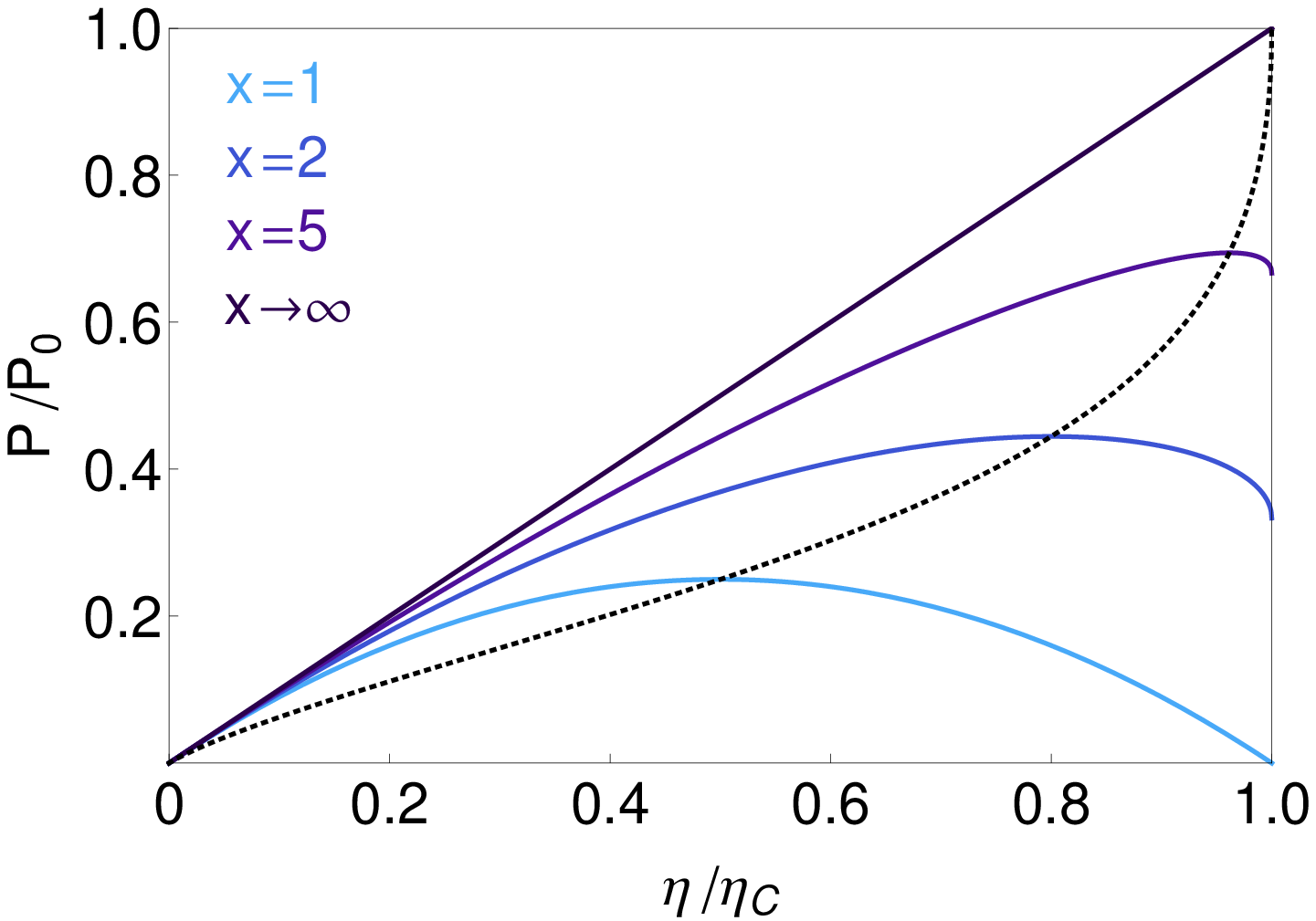,scale=0.57}
\epsfig{file=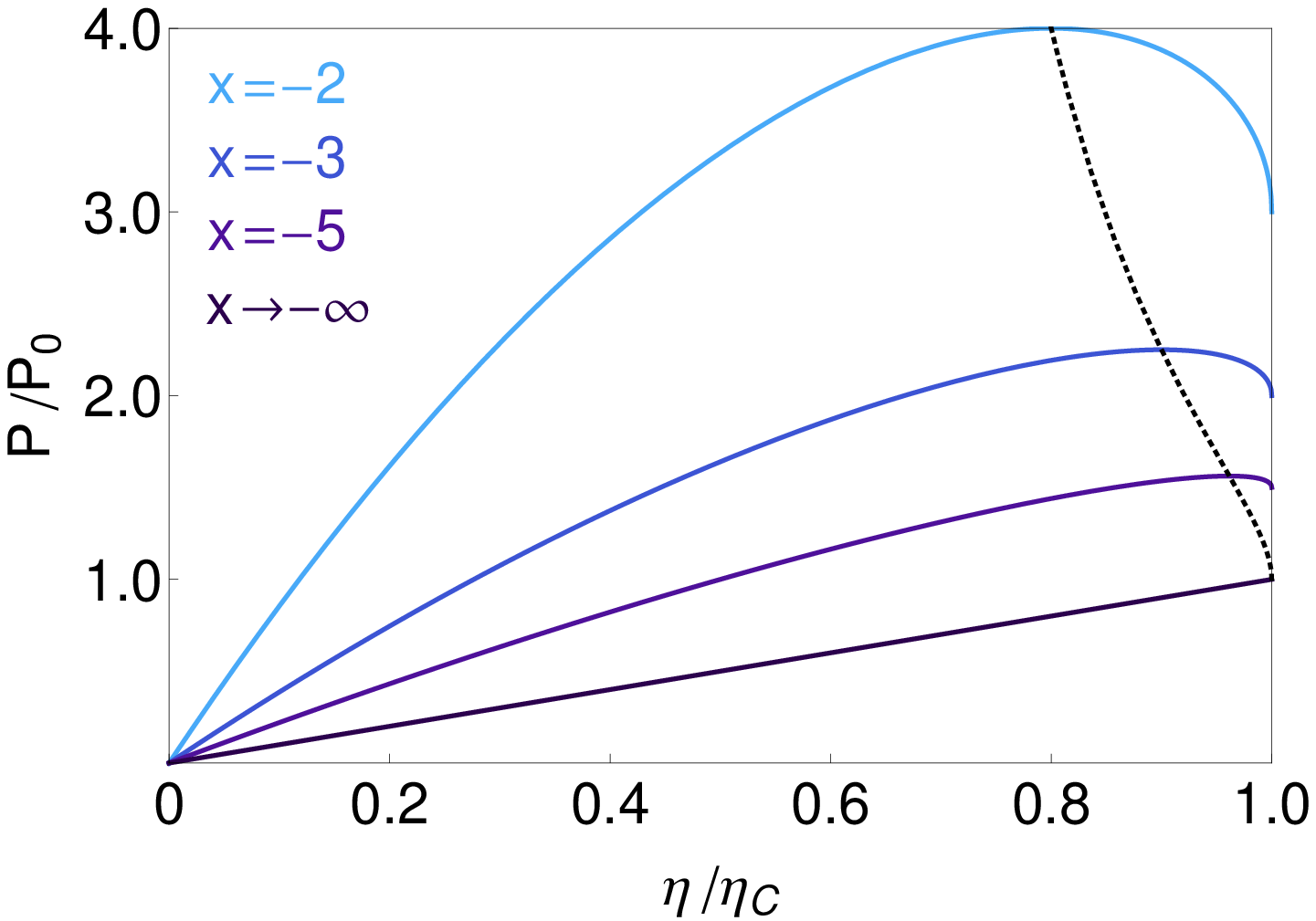,scale=0.57}
\caption{Plots of the maximum power \eqref{BPE_MaxPowEta} in units of 
$P_0\equiv T_{{\rm c}}\F_q^2L_{qq}$ as a function of the 
normalized efficiency $\bt\equiv\eta/\eta_{{\rm C}}$ for selected
values of the asymmetry parameter $x\geq 1$ in the upper and $x<-1$
in the lower panel.
For $x\rightarrow\pm\infty$, convergence to $P/P_0= \bt$ is observed.
The dotted black lines correspond to the maxima of 
\eqref{BPE_MaxPowEta} for $0\leq x < \infty$ in the first and $-\infty
< x < -1$ in the second plot thus indicating the relation between
maximum power and efficiency at maximum power.
The apparent divergence for negative $x$ occurs in the limit
$x\rightarrow-1$ \cite{Note2}.
\label{Fig_BPE}
}
\end{figure}

For a quantitative assessment of the relation between power and 
efficiency, following \cite{Brandner2015}, we consider the maximum
power at given efficiency 
\begin{multline}\label{BPE_PowEffRel}
P(\eta,x,y) =  T_{{\rm c}}\F_q^2L_{qq}\bt
                  \left(\frac{x(2+y)-y\bt}{2x(1+y)}\right.\\
                  \left.
                 +\sqrt{\frac{y^2(x+\bt)^2}{4x^2(1+y)^2}
                 -\frac{y\bt}{x(1+y)}}\;\right),
\end{multline}
as a joint benchmark parameter, which is found by eliminating $\F_w$
in \eqref{BPE_Power} in favor of 
\begin{equation}\label{BPE_RnormEtaDef}
\bt\equiv\eta/\eta_{{\rm C}}
\end{equation}
using \eqref{BPE_EfficiencyLin}.
To ensure that the Carnot value $\eta_{{\rm C}}$ is included in the 
range of accessible efficiencies, $y$ must be replaced by its bound 
$h(x)$ in \eqref{BPE_PowEffRel}. 
The resulting function 
\begin{equation}\label{BPE_MaxPowEta}
P(\eta,x)\equiv P(\eta,x,h(x))
\end{equation}
is plotted in \figref{Fig_BPE}. 
While $P(\eta,x)$ reduces to 
\begin{equation}\label{BPE_PowSym}
P(\eta,1)=T_{{\rm c}}\F_q^2L_{qq}\bt(1-\bt)
\end{equation}
in the symmetric case $x=1$ and thus vanishes linearly in the 
limit $\eta\rightarrow\eta_{{\rm C}}$, strikingly, we observe that
Carnot efficiency might be reached at finite power 
\footnote{The function \eqref{BPE_PowCEff} diverges as $x$ approaches
$-1$ from below. 
The following argument, however, shows that this singularity is only 
apparent. 
In order to obtain \eqref{BPE_PowEffRel}, we have fixed $\eta$ by 
setting
\begin{equation*}
\F_w = -\F_q \frac{L_{qq}}{L_{wq}}\Biggl(\frac{(\bt+x)y}{2(1+y)}
-\sqrt{\left(\frac{(\bt+x)y}{2(1+y)}\right)^2-\frac{\bt xy}{1+y}}
\Biggr).
\end{equation*}
For $y=h(x)$, $x<-1$ and $\eta=\eta_{{\rm C}}$, this expression
reduces to 
\begin{equation*}
\F_w = \F_q \frac{L_{qq}}{L_{wq}}\frac{2x}{x+1}.
\end{equation*}
Consequently, due to the linear response assumption $|\F_w|\ll
\bar{E}/T_{{\rm c}}$, \eqref{BPE_PowCEff} is only valid for 
\begin{equation*}
\F_q L_{qq}\ll \frac{\bar{E}}{T_{{\rm c}}}|L_{wq}|\frac{x+1}{2x},
\end{equation*}
where $\bar{E}$ denotes the typical energy scale of the unperturbed 
system. 
}
\begin{equation}\label{BPE_PowCEff}
P(\eta_{{\rm C}},x) = T_{{\rm c}}\F_q^2 L_{qq}
\frac{x-1}{x+1}
\quad\text{for}\quad |x|>1.
\end{equation}
This result is \emph{a priori} surprising, since this analysis has
fully incorporated the constraints imposed by the second law. 

\subsection{A New Constraint}\label{Sec_SHEBoundsPowEff}

We will now prove the existence of an additional constraint on the 
kinetic coefficients \eqref{LSH_OCDef}, which, so far, has been 
missing in our considerations. 
To this end, we introduce the symmetric matrix 
\begin{equation}\label{BPE_A}
\mathbb{A}\equiv\left(\!\begin{array}{ccc}
N_{qq} & L_{qw} & L_{qq}\\
L_{qw} & L_{ww} & \frac{1}{2}(L_{wq}+L_{qw})\\
L_{qq} & \frac{1}{2}(L_{wq}+L_{qw}) & L_{qq}
\end{array}\!\right),
\end{equation}
where 
\begin{equation}\label{BPE_Normalization}
N_{qq} \equiv -\frac{1}{k_B}\ps{\d g_q\L^{0\dagger} \d g_q}
\end{equation}
will play the role of a normalization constant and $\ps{\bullet}$ was
defined in \eqref{LSH_AvDef}.
The matrix $\mathbb{A}$ has the nontrivial property of being positive
semidefinite such that the determinant of any of its principal
submatrices must be nonnegative, as we show in appendix \ref{Apx_A} by
using only the rather general assumptions of \secref{\ref{Sec_OC}}.

Two important implications follow immediately from this insight. 
First, by taking the determinant of the lower right $2\times 2$-
submatrix, we recover the inequality \eqref{LSH_SecondLaw}, which we
inferred from the second law on the phenomenological level and now
have proven explicitly. 
Second, by evaluating the determinant of $\mathbb{A}$, we get the new
constraint
\begin{align}
L_{qq} &  \leq N_{qq} \frac{L_{ww}L_{qq}-(L_{wq}+L_{qw})^2/4}
{L_{ww}L_{qq}-L_{wq}L_{qw}},\label{BPE_LhyBoundExpl}\\[3pt]
       &  = N_{qq}\left(1-y/h(x)\right), \label{BPE_LhyBound}         
\end{align}
which, in contrast to the bare second law, leads to a bound on
power.
Specifically, this bound is found by bounding $L_{qq}$ in 
\eqref{BPE_PowEffRel} using \eqref{BPE_LhyBound} and then maximizing
the resulting function with respect to $y$ \cite{Brandner2015}.
This procedure yields the simple result
\begin{equation}\label{BPE_PowerBoundNew}
P(\eta,x)\leq 
\hat{P}(\eta,x)\equiv 4 \bar{P}_0 \begin{cases}
\bt(1-\bt)    & \text{for}\quad |x|\geq1 \\
\bt-\bt^2/x^2 & \text{for}\quad |x|<1
\end{cases},
\end{equation}
where we define the standard power 
\begin{equation}
\bar{P}_0\equiv T_{{\rm c}}\F_q^2N_{qq}/4
\end{equation}
and $y$ becomes
\begin{align}
y^\ast(x,\eta) & = \frac{4x\bt}{(x-1)(1+x-2\bt)}\quad & &\text{for}\quad
|x|\geq 1,\nonumber\\
y^\ast(x,\eta) & = \frac{4\bt}{x-x^3-2\bt+2x\bt}\quad & &\text{for}\quad
|x|<1
\end{align}
through the optimization. 
Remarkably, for any $|x|\geq 1$ the bound \eqref{BPE_PowerBoundNew}, 
which is our third main result, restores the quadratic relation 
between power and efficiency \eqref{BPE_PowSym}, which we found in our
first analysis not invoking the new bound \eqref{BPE_LhyBound} only 
for the symmetric case $x=1$. 
Consequently, we have shown that, in the linear response regime,
the power of any cyclic heat engine comprised by our theoretical
framework must vanish at least linearly as its efficiency approaches
the Carnot value. 
We emphasize that this quite natural result can neither be derived 
from the laws of thermodynamics nor from micro-reversibility, which 
appears in form of the reciprocity relation \eqref{LSH_RcpRel}.
Instead, it relies on the additional constraint
\eqref{BPE_LhyBoundExpl}, which is beyond both of these principles. 

\section{An Illustrative Example}\label{Sec_Example}
\subsection{Model and Kinetic Coefficients}
\begin{figure}
\epsfig{file=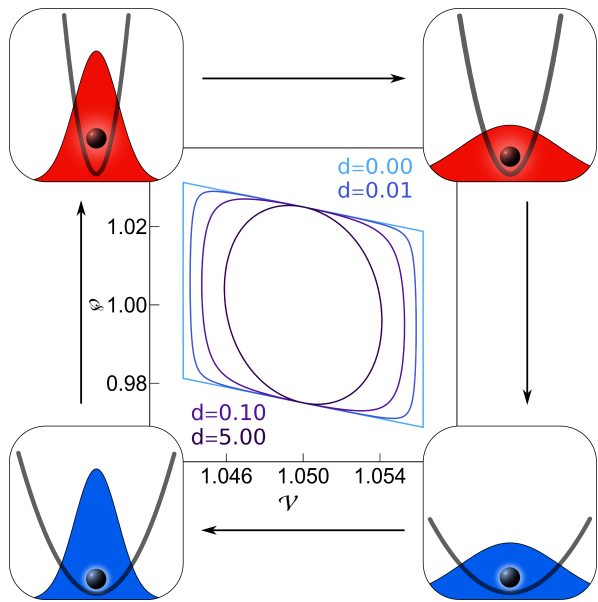,scale=0.185}
\caption{\label{Fig_SHE}
Operation cycle of a Brownian heat engine.
The vertical axis corresponds to the normalized time-dependent 
strength of the harmonic trap in units of $\wp\equiv\k(t)/\k_0$, the
horizontal axis to the normalized width $\mathcal{V}\equiv \langle x^2
\rangle/(2 x_0^2)$ of the distribution function. 
This plot is analogous to the pressure volume diagram of a macroscopic
heat engine such that the area encircled by the colored lines 
quantifies the work  extracted per operation cycle. 
Specifically, the plots were obtained using the protocols 
\eqref{Ex_OptProt} and \eqref{Ex_IntPolProt} for different values of
the shape parameter $d$, $2\mu\k_0\T=1$, $\eta_{{\rm C}}=
T_{{\rm c}}\F_q =1/10$ and $\bt=1/2$.
The small graphics show sketches of the potential (gray line) and the
the phase space distribution, whose color reflects the temperature of 
the heat bath, at the respective edges of the cycle. 
For further explanations, see \secref{\ref{Sec_Example}}.}
\end{figure}
A particularly simple setup for a stochastic heat engine consists of
a Brownian particle in one spatial dimension confined in a harmonic 
potential of variable strength $\k(t)$ and immersed in a heat bath of
time-dependent temperature $T(t)$ as schematically shown in 
\figref{Fig_SHE}.
Originally proposed in \cite{Schmiedl2008}, this model has recently
been realized in a remarkable experiment \cite{Blickle2011} and can be
used to illustrate various aspects of stochastic thermodynamics like 
the role of feedback \cite{Abreu2011,Bauer2012} and shortcuts to 
adiabaticity \cite{Tu2014}. 

Here, by applying our general theory developed in the last sections
we will calculate the kinetic coefficients for this stochastic heat 
engine and optimize the protocol $\k(t)$ controlling the trap 
strength to obtain maximum power for given efficiency.

In the overdamped limit, due to the absence of kinetic energy, the 
Hamiltonian of the system
\begin{align}
& H(x,t) = H_0(x) + \D H g_w(x,t) \quad\text{with}\nonumber\\
& H_0(x)\equiv \frac{\k_0}{2}x^2, \quad
\Delta H\equiv \k x_0^2, \quad
g_w(x,t)\equiv  \frac{x^2}{2x_0^2}\gt_w(t)
\end{align}
depends only on the position $x$ of the particle. 
Here, $\k_0$ is the equilibrium strength of the trap, $\k$ 
quantifies the strength of the time-dependent driving, $\gt_w(t)$ 
denotes the driving protocol and $x_0\equiv\sqrt{2k_BT_{{\rm c}}
/\k_0}$ is the characteristic length scale of the system. 
The time evolution of the probability density $p(x,t)$ for finding 
the particle at the position $x$ is generated by the Fokker-Planck 
operator
\begin{equation}\label{Ex_FPE}
\L(t)\equiv \mu \left(\k_0 + \k g_w(t)\right)
\partial_x x  + \mu k_B T(t)\partial_x^2,
\end{equation}
which, in equilibrium reduces to 
\begin{equation}\label{Ex_PFEEq}
\L^0\equiv \mu\k_0 \partial_x x + \mu k_B T_{{\rm c}}\partial_x^2.
\end{equation}
Here, $\mu$ denotes the mobility and the temperature $T(t)\equiv 
T_{{\rm c}}T_{{\rm h}}/(T_{{\rm h}}-\D T \gt_q(t))$ oscillates between
the cold and the hot levels $T_{{\rm c}}$ and $T_{{\rm h}}\equiv 
T_{{\rm c}}+\D T$. 
The equilibrium fluctuations showing up in \eqref{LSH_OCJJCorr} read
\begin{equation}
\d g_\alpha(x,t)  = \g_\alpha(t)\k_0\xi_\alpha 
                  \left(x^2-k_BT_{{\rm c}}/\kappa_0\right)
\end{equation}
with
\begin{equation}
\xi_w\equiv 1/(4k_BT_{{\rm c}})\quad\text{and}\quad \xi_q\equiv -1/2.
\end{equation}
Formula \eqref{LSH_OCJJCorr} can be easily evaluated by using the 
detailed balance relation \eqref{LSH_DB} to transform $\L^0$ into
$\L^{0\dagger}$, since it is readily seen that the function 
$x^2-k_BT_{{\rm c}}/\kappa_0$ is a right eigenvector of 
$\L^{0\dagger}$ with corresponding eigenvalue $-2\mu\kappa_0$.
The resulting kinetic coefficients
\begin{multline}\label{Ex_OCFunctl}
L_{\alpha\beta} = -\frac{2k_BT_{{\rm c}}^2\xi_\alpha\xi_\beta}{\T}
                  \tint\biggl(\dot{\gt}_\alpha(t)\gt_\beta(t)\\ 
                  -\tauint \dot{\gt}_\alpha(t)
                  \dot{\gt}_\beta(t-\tau) e^{-2\mu\k_0\tau}\biggr)
\end{multline}
are functionals of the protocols $\gt_\alpha(t)$. 
Note that, besides the general reciprocity relation 
\eqref{LSH_RcpRel}, these coefficients also satisfy  the special
symmetry relation \eqref{LSH_SpRcpRel}, since the factorization 
condition \eqref{LSH_gFctzn} is fulfilled in the example discussed
here. 
\begin{figure}
\flushleft
\epsfig{file=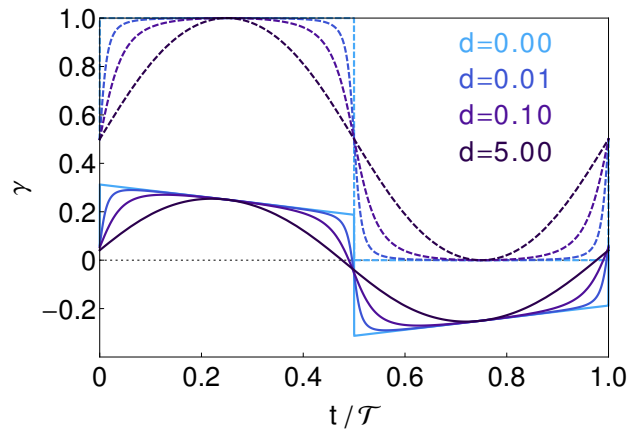,scale=0.57}
\caption{Plots of the temperature protocol $\g_q(t,d)$ defined in
\eqref{Ex_IntPolProt} (dashed lines) and the corresponding
optimal protocol $\g_w^\ast(t,\eta,d)$ for the trap
strength (solid lines) obtained from \eqref{Ex_MaxPowProt} for four
different values of the parameter $d$ as functions of $t/\T$
\cite{Note3}.
For all plots, we have set $2\mu\k_0\T=1$ and $\bt=1/2$. 
Additionally, the optimal protocol $\g_w^\ast(t)$ has been rescaled by 
the dimensionless factor $\k_0\eta_{{\rm C}}/\k$.
\label{Fig_Protocols}}
\end{figure}

\subsection{Optimization}

The optimal protocol $\gt^\ast_w(t,\eta)$ for the strength of
the harmonic trap for a given time dependence of temperature $\gt_q(t)$
maximizes the power output at fixed efficiency $\eta$. 
It is determined by the variational condition 
\begin{equation}\label{Ex_OptProt}
0\overset{!}{=}\left.\frac{\delta}{\delta\gt_w(t)}P[\gt_w(t), \gt_q(t)]
\right|_{P/J_q\overset{!}{=}\eta},
\end{equation}
where the power $P$ and the heat flux $J_q$ are regarded as
functionals of $\gt_w(t)$ and $\gt_q(t)$.
As we show in appendix \ref{Apx_OptProt}, this constrained 
optimization problem has the general solution
\begin{multline}\label{Ex_MaxPowProt}
\g_w^\ast(t,\eta) =\frac{\k_0\eta_{{\rm C}}}{\k}\Bigl((1-\bt)\g_q(t)\\
-2\bt\mu\k_0\int_0^t \!\!\! d\tau \left(\g_q(\tau)-\bar{\g_q}\right)
\Bigr)+\g_0
\end{multline}
for $0\leq t\leq \T$, where we used the abbreviations
\eqref{BPE_RnormEtaDef},
\begin{equation}\label{Ex_AbbrevDef}
\bar{\g}_q \equiv \frac{1}{\T}\tint \g_q(t)
\end{equation}
and $\g_0$ denotes an arbitrary constant.
Using this protocol, the maximum power 
\begin{equation}\label{Ex_MaxPow}
P_{{\rm max}}(\eta)=
       \frac{k_BT_{{\rm c}}^3 \F_q^2\mu\k_0}{\T}\bt(1-\bt)\tint 
       \bigl(\g_q(t)- \bar{\g}_q\bigr)^2,
\end{equation}
can be extracted per operation cycle at efficiency $\eta$.

\subsection{Comparison with General Bound}
 
In order to compare this result with the general bound
\eqref{BPE_PowerBoundNew}, we evaluate the normalization constant
\begin{equation}
N_{qq}=\frac{k_B T_{{\rm c}}^2\mu\k_0}{\T}\tint \g^2_q(t)
\end{equation}
defined in \eqref{BPE_Normalization} and rewrite \eqref{Ex_MaxPow} as
\begin{equation}
P_{{\rm max}}(\eta) = 4\psi \bar{P}_0\bt(1-\bt),
\end{equation}
where $\bar{P}_0$ is the standard power introduced in
\eqref{BPE_PowerBoundNew}.
The dimensionless factor 
\begin{equation}\label{Ex_PsiIneq}
0\leq \psi\equiv\frac{\tint\left(\g_q(t)-\bar{\g}_q\right)^2}
{\tint \g_q^2(t)}\leq 1
\end{equation}
quantifies how close the maximum power found in the optimization comes
to the general bound \eqref{BPE_PowerBoundNew}.
Since, $0\leq\g_q(t)\leq 1$, it is reached only for $\bar{\g}_q
\rightarrow 0$.
This limit, however, requires $\g_q(t)\rightarrow 0$ for any $t$ and
thus, inevitably, leads to vanishing absolute power. 

For an illustration of this issue, we chose $\g_q(t)$ as the step
function \eqref{LSH_StepProt} such that the system is alternately in
contact with a hot and a cold bath, respectively during the time
intervals $\T_1$ and $\T-\T_1$.
We then find $\psi=1-\T_1/\T$ and $P_0\sim \T_1/\T$.
Thus, as $\T_1$ is decreased, $\psi$ comes arbitrarily close to $1$
and $P_0$ decays linearly to zero. 
This example shows that our bound is asymptotically tight.

A particular advantage of our approach is that it allows to treat
situations with a continuously varying temperature of the environment
on equal footing with the scenario proposed in \cite{Schmiedl2008},
which involves instantaneous switchings between a hot and a cold
reservoir. 
In order to illustrate this feature, we consider the specific choice
\begin{equation}\label{Ex_IntPolProt}
\g_q(t,d)\equiv \frac{\sqrt{1+d}\sin[2\pi t/\T] }
                    {2\sqrt{\sin^2[2\pi t/\T] +d}}+\frac{1}{2}
\end{equation}
for the protocol $\g_q(t)$, which, in the linear response regime, is
proportional to the temperature $T(t)$.
The function \eqref{Ex_IntPolProt}, which interpolates between a step
function $(d\rightarrow 0)$ and a simple sine $(d\rightarrow
\infty)$ \cite{Berger2009}, is plotted together with the corresponding
optimal protocol $\g_w^\ast(t,\eta,d)$
\footnote{Here, we have chosen the constant $\g_0$ showing up in 
\eqref{Ex_OptProt} as
\begin{equation}
\g_0 = \frac{\k_0\eta_{{\rm C}}}{2\k}\biggl(1-\bt
-\frac{2\mu\k_0\T}{\pi}\sqrt{1+d}\cdot
 \arctan\left[ 1/\sqrt{d}\right] \biggr)
\end{equation}
such that 
\begin{equation*}
\tint \g_w^\ast(t,\eta,d) = 0,
\end{equation*}
ensuring that the constant part of the potential is included in
$H_0(\x)$.}
in \figref{Fig_Protocols}. 
We find that, for $d=0$, this protocol shows two sudden jumps
occurring simultaneously with the instantaneous changes of the 
bath temperature.
Such discontinuities were shown to be typical for thermodynamically
optimized finite-time protocols connecting two equilibrium states 
\cite{schm07}.
Since, here, we are concerned with periodic states generated by
permanent driving rather than a transient process with equilibrium
boundary conditions, it is, however, not surprising that both, the
temperature and the trap strength protocol, become continuous as the
shape parameter $d$ is increased and, respectively, converge to two
sines in phase in the limit $d\rightarrow\infty$.

For the protocols \eqref{Ex_OptProt} and \eqref{Ex_IntPolProt}, 
the parameter $\psi$ becomes 
\begin{equation}
\psi = \frac{1+d}{2(1+d) +\sqrt{d(1+d)}}.
\end{equation}
This function decays monotonically from $1/2$ for $d=0$ to $1/3$ for
$d\rightarrow\infty$. 
Consequently, $\psi$ can not reach its maximum $1$ within the class of
protocols \eqref{Ex_IntPolProt}.
This limitation can be understood from the argument given below
\eqref{Ex_PsiIneq}, since, for any $d$, we have $\bar{\g}_q=1/2$.

The standard power $P_0$ is proportional to the function 
$2+d-\sqrt{d(1+d)}$, which decays monotonically from $2$ to 
$3/2$ as $d$ increases from $0$ to infinity. 
Thus, $P_0$ exhibits the same qualitative dependence on the shape
parameter $d$ as the efficiency. 
We can therefore conclude that, at least within the model considered
here, a steeply rising and falling temperature performs better than a
smoothly changing one. 

\section{Concluding Perspectives}

In this work, we have demonstrated that non-equilibrium periodic
states, which emerge naturally in periodically driven systems, can 
be endowed with the universal structure of irreversible 
thermodynamics. 
Moreover, by using a quite general stochastic approach, we have 
proven the generalized reciprocity relation \eqref{LSH_RcpRel} for the
kinetic coefficients characterizing the linear response regime. 

Our new framework is particularly useful for a systematic study of
the performance of cyclic heat engines. 
Within the linear regime, bounding the power of these machines in such
a way that the rather peculiar option of Carnot efficiency at finite
power is ruled out requires the new relation \eqref{BPE_LhyBoundExpl}.
This constraint is beyond the laws of thermodynamics and the principle
of microscopic reversibility and has been proven here for the first
time on a general level. 
Remarkably, up to the normalization factor, an identical bound has
been discovered only recently in a numerical analysis of a particular
class of thermoelectric heat engines \cite{Brandner2015}. 
Whether this intriguing similarity suggests the existence of a so far
undiscovered universal principle that applies to periodic as
well as to steady states leading to a bound on power for any type of 
heat engine that operates in the linear regime remains an exciting
topic for future research. 

A promising starting point for investigations in this directions might
be found in the Green-Kubo relations, which follow from first
principles and provide general expressions for the conventional 
kinetic coefficients in terms of equilibrium correlation functions
\cite{Kubo1998}.
Using a Fokker-Planck approach, we have shown that an analogous
representation for the kinetic coefficients exists in periodically
driven systems. 
The quantities related by the relevant correlation functions are,
however, well defined irrespective of specific dynamics governing the
time evolution of the phase space density. 
It might therefore be possible to obtain Hamiltonian-based expressions
also for the periodic kinetic coefficients introduced in this work.
Finding a proper way to take the time dependence of temperature into
account is arguably the major challenge here. 

This problem also prevents an immediate extension of our formalism
to the quantum realm.
While the first part of our analysis, the identification of proper
fluxes and affinities, carries over to quantum mechanics line by line,
it is not clear at the moment whether the constraints on the kinetic
coefficients obtained here classically can be likewise transferred or
properly generalized.
This topic appears all the more urgent in the light of recent 
developments \cite{Abah2012,an15,Pekola2015} showing that the emerging
field of quantum thermodynamics nowadays comes within the range of
experiments.

\appendix

\section{Derivation of the Expression \eqref{LSH_OCJJCorr} for the 
Onsager Coefficients}\label{Apx_DevOC}

For an expression of the kinetic coefficients \eqref{LSH_OCDef} 
depending only on equilibrium quantities the perturbations $\L^X(t)$
showing up in the linear response solution \eqref{LSH_LinSol} have to
be eliminated. 
To this end, we invoke the property \eqref{LSH_InstEq} of the full 
Fokker-Planck operator. 
Substituting \eqref{LSH_ExpFPO} into \eqref{LSH_InstEq}, expanding 
the exponential in $\Delta H$ and $\Delta T$ and collecting linear
order terms provides us with the relations
\begin{equation}\label{LSH_PertIdentities}
\begin{split}
& \Delta H \L^H(t)\pq  = \F_w\L^0 g_w(\x,t)\pq/k_B,\\
& \Delta T \L^T(t)\pq  = \F_q\L^0 g_q(\x,t)\pq/k_B,
\end{split}
\end{equation}
where we used the definition \eqref{SHE_gqDef}.
Up to corrections of order $\Delta^2$, the periodic distribution 
$\plc$ can thus be rewritten as 
\begin{widetext}
\begin{align}\label{ADOC_LinSolF}
\plc & = \pq 
       + \sum_{\alpha=w,q}\frac{\F_\alpha}{k_B}\tauint e^{\L^0\tau}
         \L^0 g_\alpha(\x,t-\tau)\pq\\
     & = \pq 
       + \sum_{\alpha=w,q}\frac{\F_\alpha}{k_B}\tauint e^{\L^0\tau}
         \L^0 \d g_\alpha(\x,t-\tau)\pq\\
     & = \pq
       -\sum_{\alpha=w,q}\frac{\F_\alpha}{k_B}
       \left(\d g_\alpha(\x,t)\pq
       -\tauint e^{\L^0\tau} \d\dot{g}_\alpha(\x,t-\tau)\pq\right).
\end{align}
\end{widetext}
In the second line, we replaced $g_\alpha(\x,t)$ with its equilibrium
fluctuation defined in \eqref{LSH_FluctDef}.
This modification does not alter the right hand side of 
\eqref{ADOC_LinSolF}, since $\L^0\pq=0$. 
However, it ensures that the function, on which the exponential
operator acts in the third line, which was obtained by an integration
by parts with respect to $t$, has no overlap with the nullspace of
$\L^0$ and thus the integral with infinite upper bound is well defined.
Note that the upper boundary term vanishes, since the operator $\L^0$
is nonpositive \cite{Risken1996}.
Inserting \eqref{ADOC_LinSolF} into \eqref{SHE_WorkFluxDef} and
\eqref{SHE_HeatFluxMeanRep} 
yields
\begin{equation}
L_{\alpha\beta}=-\frac{1}{k_B}\ps{\dot{g}_\alpha \d g_\beta}
+\frac{1}{k_B}\tauint\ps{\dot{g}_\alpha(0);\d\dot{g}_\beta(-\tau)}.
\end{equation}
Herein, obviously, $\dot{g}_\alpha(\x,t)$ can be replaced by its
equilibrium fluctuation $\d\dot{g}_\alpha(\x,t)$ in the first term. 
Since any constant lies in the left nullspace of $\L^0$ and, hence, is
orthogonal to the function $e^{\L^0\tau}\d\dot{g}_\alpha(\x,t-\tau)\pq
$, the same replacement can be carried out in the second term without
leading to additional contributions such that \eqref{LSH_OCJJCorr} is,
finally, obtained.

\section{Reciprocity Relations}\label{Apx_RcpRel}
In this appendix, we establish the reciprocity relations 
\eqref{LSH_RcpRel} and \eqref{LSH_SpRcpRel}.
To this end, we first recall formula \eqref{LSH_OCJJCorr}, which 
becomes
\begin{align}
& L_{\alpha\beta}[H(\x,t),T(t),\B]\nonumber\\
& = -\frac{1}{k_B\T}\txint\biggl(
	 \dg_\alpha(\x,t) \d g_\beta(\x,t)\pq\nonumber\\
& \quad\; + \tauint \dg_\alpha(\x,t)e^{\L^0\tau}
                    \dg_\beta(\x,t-\tau)\pq\biggr),
\label{ADRR_GreenKubo}
\end{align}
using the definitions \eqref{LSH_AvDef} and \eqref{LSH_EqCorr}.
By applying the detailed balance relation \eqref{LSH_DB} and changing
the integration variable $t$ according to $t\rightarrow \T-t$, this 
expression can be transformed to 
\begin{align}
& L_{\alpha\beta}[H(\x,t),T(t),\B]\nonumber\\
& = -\frac{1}{k_B\T}\txint\biggl(
	-\dg_\alpha(\x,-t)\d g_\beta(\x,-t)\pq\nonumber\\
& \quad\; + \tauint \dg_\beta(\x,-t)e^{\tL^0\tau}
                    \dg_\alpha(\x,\tau-t)\pq\biggr),
\label{ADRR_GreenKuboDB}
\end{align}
where we introduced the shorthand notation $\tL^0\equiv\L^0(\be\x)$.
Furthermore, to transfer the variable $\tau$ from the argument of 
$\dg_\beta$ in \eqref{ADRR_GreenKubo} to the argument of
$\dg_\alpha$ in \eqref{ADRR_GreenKuboDB}, we used the identity 
\begin{align}
&\tint a(t)b(t-\tau) = \int_{-\tau}^{\T-\tau} \!\!\! dt \; a(t+\tau)b(t)\nonumber\\
&=\tint a(t+\tau)b(t) +\int_{-\tau}^0 \!\!\! dt \; a(t+\tau)b(t)\nonumber\\
&                     -\int_{\T-\tau}^\T \!\!\! dt \; a(t+\tau)b(t)
=\tint a(t+\tau)b(t),
\end{align}
which holds for any two $\T$-periodic functions $a(t)$ and $b(t)$.
Finally, we reverse the magnetic field as well as the driving 
protocols and apply the change of integration variables 
$\x\rightarrow\be\x$, whose Jacobian is $1$. 
By exploiting the symmetry $\d g_\alpha(\be\x,t)=\d g_\alpha(\x,t)$,
which follows from condition \eqref{LSH_HamiltonianTRS}, and carrying
out one integration by parts with respect to $t$ in the first summand,
we obtain
\begin{align}
& L_{\alpha\beta}[H(\x,-t),T(-t),-\B]\nonumber\\
& = -\frac{1}{k_B\T}\txint\biggl(
	 \dg_\beta(\x,t)\d g_\alpha(\x,t)\pq\nonumber\\
& \quad\; + \tauint
\dg_\beta(\x,t)e^{\L^0\tau}\dg_\alpha(\x,t-\tau)\pq\biggr)\\
& =L_{\beta\alpha}[H(\x,t),T(t),\B]
\end{align}
thus completing the proof of the reciprocity relation 
\eqref{LSH_RcpRel}.

We now turn to the special case where the function $g_w(\x,t)$ can be
separated in the form
\begin{equation}
g_w(\x,t) = g_w(\x)\gt_w(t).
\end{equation}
Plugging this expression into \eqref{ADRR_GreenKubo} and invoking the 
definition $g_q(\x)\equiv-H_0(\x)$ yields the expression
\begin{align}
& L_{\alpha\beta}[\gt_w(t),\ta(t),\B]\nonumber\\
& = -\frac{1}{k_B\T}\txint\biggl(\dot{\gt}_\alpha(t)\gt_\beta(t)
	\d g_\alpha(\x)\d g_\beta(\x)\pq\nonumber\\
& + \! \tauint \dot{\gt}_\alpha(t)\dot{\gt}_\beta(t-\tau)
    \d g_\alpha(\x)e^{\L^0\tau}\d g_\beta(\x)\pq\biggr),
\label{ADRR_GreenKuboSp}
\end{align}
which, by virtue of the detailed balance relation \eqref{LSH_DB},
equals
\begin{align}
& L_{\alpha\beta}[\gt_w(t),\gt_q(t),\B]\nonumber\\
& = -\frac{1}{k_B\T}\txint\biggl(\dot{\gt}_\alpha(t)\gt_\beta(t)
	\d g_\alpha(\x)\d g_\beta(\x)\pq\nonumber\\
& + \! \tauint \dot{\gt}_\alpha(t)\dot{\gt}_\beta(t-\tau)
    \d g_\beta(\x)e^{\tL^0\tau}\d g_\alpha(\x)\pq\biggr).
\label{ADRR_GreenKuboSpDB}
\end{align}
Relation \eqref{LSH_SpRcpRel} can now be obtained by following the 
same steps as in the general case, that is, by applying the 
transformation $\x\rightarrow\be\x$, reversing the magnetic field and
interchanging the arguments $\gt_w(t)$ and $\gt_q(t)$ of 
$L_{\alpha\beta}$, using the symmetry $\d g_\alpha(\x)
=\d g_\alpha(\be\x)$ and performing one integration by parts in the
first term.

\section{Positivity of $\mathbb{A}$}\label{Apx_A}

The proof that the matrix $\mathbb{A}$ defined in \eqref{BPE_A} is 
positive semidefinite consists of two major steps. 
First, for arbitrary numbers $y_w,y_q\in\mathbb{R}$, we consider the
quadratic form 
\begin{align}
& \mathcal{Q}_0(y_w,y_q) 
	\equiv\sum_{\alpha,\beta=w,q} L_{\alpha\beta}y_\alpha y_\beta,\\
&= \frac{1}{k_B}\tauint\ps{G(0);G(-\tau)}\nonumber\\
&= \frac{1}{k_B\T}\ttauint \ev{G(t)e^{\tK\tau}G(t-\tau)}\nonumber
\end{align}
where we used the expression \eqref{LSH_OCJJCorr} for the kinetic 
coefficients, defined
\begin{equation}
G(\x,t)\equiv\sum_{\alpha=w,q} y_\alpha \dg_\alpha(\x,t)
\end{equation}
and applied the detailed balance relation \eqref{LSH_DB} to obtain the
second line. 
We recall the definitions \eqref{LSH_AvDef} for the meaning of the 
angular brackets. 
The crucial ingredient for this first step consists of the identity
\begin{widetext}
\begin{align}
& -\frac{1}{2k_B\T}\tint
    \ev{\tauint e^{\tK\tau} G(t-\tau) \bigl(\K + \tK\bigr)\!
    \taupint e^{\tK\tau'} G(t-\tau')}\label{AA_QOProd0}\\
& = -\frac{1}{2k_B\T}\tint\!\!\!\tauint\!\!\!\!\taupint\Biggl\{
    \ev{\Bigl(\bigl(\partial_\tau e^{\tK\tau}\bigr)G(t-\tau)\Bigr)
                                  e^{\tK\tau'}G(t-\tau')}
    +\ev{\Bigl( e^{\tK\tau}G(t-\tau)\Bigr)
                                  \bigl(\partial_{\tau'}e^{\tK\tau'}\bigr)
                                  G(t-\tau')}
    \Biggr\}\label{AA_QOProd1}\\
& = \frac{1}{2k_B\T}\ttauint \Biggl\{
    2\ev{G(t)e^{\tK\tau}G(t-\tau)}
    +\partial_t \taupint\ev{\Bigl( e^{\tK\tau}G(t-\tau)\Bigr)
    e^{\tK\tau'}G(t-\tau')}
    \Biggr\}\label{AA_QOProd2}\\
& = \frac{1}{k_B\T}\ttauint\ev{G(t)e^{\tK\tau}G(t-\tau)}
  = \mathcal{Q}_0(y_w,y_q).
\end{align}
\end{widetext}
Here, we used the relation 
\begin{equation}\label{AA_Kadj}
\ev{A\K B} = \ev{B \tK A},
\end{equation}
which holds for any functions $A(\x), B(\x)$ by virtue of the detailed
balance condition \eqref{LSH_DB}, to obtain \eqref{AA_QOProd1} from 
\eqref{AA_QOProd0}.
Expression \eqref{AA_QOProd2} follows by applying an integration by
parts with respect to $\tau$ and $\tau'$, respectively, in the first
and the second summand of \eqref{AA_QOProd1}.
Finally, the second contribution in \eqref{AA_QOProd2} vanishes after
carrying out the $t$-integration, since the function $G(\x,t)$ 
is $\T$-periodic in time. 

Next, we note that \eqref{LSH_DB} implies
\begin{align}
&\ev{A\bigl(\K+\tK\bigr)A}\label{AA_NegAv}\\
&= \xint A(\x)\bigl(\L^{0}\pq + \pq\K\bigr)A(\x)\nonumber\\
&= \xint [\pq]^\frac{1}{2}A(\x)
   \bigl(\mathsf{K}^0+ \mathsf{K}^{0\dagger}\bigr)[\pq]^\frac{1}{2}A(\x),
         \nonumber
\end{align}
where 
\begin{equation}
\mathsf{K}^0\equiv [\pq]^{-\frac{1}{2}}\L^0 [\pq]^\frac{1}{2}.
\end{equation}
Since the Hermitian part of this operator is negative semidefinite 
\cite{Risken1996}, it follows that \eqref{AA_NegAv} is non-positive
for any $A(\x)$. 
Hence, we can conclude that the quadratic form 
$\mathcal{Q}_0(y_w,y_q)$ is positive semidefinite, since it can be 
written in the form \eqref{AA_QOProd0}.

For the second step of the proof, we introduce the quadratic form
\begin{equation}\label{AA_QDef}
\mathcal{Q}(y_w,y_q,z)\equiv
\mathcal{Q}_0(y_w,y_q)+\mathcal{Q}_1(y_w,y_q,z)
\end{equation}
with
\begin{align}\label{AA_Q1Def}
& \mathcal{Q}_1(y_w,y_q,z)\equiv N_{qq}z^2 
                          + 2z\sum_{\alpha=w,q}L_{q\alpha} y_\alpha\\
& = -\frac{1}{k_B}\left(\ps{F\L^{0\dagger}F}-2\ps{FG}
         -2\tauint\ps{\dot{F}(0);G(-\tau)}\right)\nonumber\\
& = -\frac{1}{2k_B\T}\tint\Biggl\{2\ev{F(t)\L^{0\dagger}F(t)}
    -4\ev{F(t)G(t)}\nonumber\\
&   \hspace*{4cm}-4\tauint\ev{\dot{F}(t)e^{\tK\tau}G(t-\tau)}\Biggr\}
    \nonumber
\end{align}
where $y_w,y_q,z\in\mathbb{R}$ and
\begin{equation}
F(\x,t)\equiv z \d g_q(\x,t).
\end{equation}
Note that, in \eqref{AA_Q1Def}, we used \eqref{LSH_OCJJCorr} and 
\eqref{BPE_Normalization} as well as the detailed balance condition
\eqref{LSH_DB}.
The expression \eqref{AA_Q1Def} can be rewritten as
\begin{align}
& \mathcal{Q}_1(y_w,y_q,z)= -\frac{1}{2k_B\T}\tint\Biggl\{
\ev{F(t)\bigl(\K+\tK\bigr)F(t)}\nonumber\\
& +\tauint\ev{F(t)\bigl(\K+\tK\bigr)
\Bigl(e^{\tK\tau} G(t-\tau)\Bigr)}\nonumber\\
& +\tauint\ev{e^{\tK\tau}G(t-\tau)\Bigr)\bigl(\K+\tK\bigr)F(t)}
\Biggr\}.
\label{AA_Q1Prod}
\end{align}
This assertion can be proven by expanding \eqref{AA_Q1Prod},
invoking \eqref{AA_Kadj} as well as the identity 
\begin{align}
\K F(\x,t) & = z \K g_q(\x,t) = -z \g_q(t)\K H_0(\x)\nonumber\\
           & =-z \g_q(t)\tK H_0(\x) = \tK F(\x,t),
\end{align}
which is implied by condition \eqref{LSH_AutDyn}, and integrating by
parts, first with respect to $\tau$ and then with respect to $t$, 
respectively, in the second and third term showing up in
\eqref{AA_Q1Prod}.
Finally, putting together \eqref{AA_QOProd0} and \eqref{AA_Q1Prod}
leads to 
\begin{widetext}
\begin{align}
& \mathcal{Q}(y_w,y_q,z) =
-\frac{1}{2k_B\T}\tint\ev{\Bigl(F(t)+\tauint e^{\tK\tau} G(t-\tau)\Bigr)
\bigl(\K+\tK\bigr)\Bigl(F(t)+\taupint e^{\tK\tau'} G(t-\tau')\Bigr)}.
\end{align}
\end{widetext}
The average showing up in this expression is of the form 
\eqref{AA_NegAv} and thus must be non positive.
Consequently, we have $\mathcal{Q}(y_w,y_q,z)\geq 0$ for any 
$y_w,y_q,z$.
Moreover, since
\begin{equation}
\mathcal{Q}(y_w,y_q,z) = \mathbf{y}^t\mathbb{A}\mathbf{y}
\end{equation}
with $\mathbf{y}\equiv (z,y_w,y_q)^t$ it follows that the matrix 
$\mathbb{A}$ must be positive semidefinite and thus the proof is 
completed. 

\section{Optimal Protocol}\label{Apx_OptProt}

The aim is to determine the optimal protocol $\g^\ast_w(t)$, which 
maximizes the rescaled output power 
\begin{equation}
\bar{P}\equiv \frac{P}{T_{{\rm c}}\F_q^2}=
-(L_{ww}\chi^2 + L_{wq}\chi)
\quad\text{with}\quad
\chi\equiv \F_w/\F_q
\end{equation}
for given time dependence of the bath temperature $g_q(t)$ and 
normalized efficiency
\begin{equation}\label{AOP_EffConstr}
\bt=-\frac{L_{ww}\chi^2 + L_{wq}\chi}{L_{qw}\chi + L_{qq}}.
\end{equation}
This task is captured by the objective functional 
\begin{align}\label{AOP_OjctFunctl}
&\mathcal{P}[\g_w(t),\g_q(t),\l]\equiv\nonumber\\
&(\l-1)\chi^2 L_{ww}+(\l-1)\chi L_{wq}+ \l\bt\chi L_{qw}+\l\bt L_{qq},
\end{align}
where the additional constraint \eqref{AOP_EffConstr} is taken into 
account by introducing the Lagrange multiplier $\lambda$. 
By inserting \eqref{Ex_OCFunctl}, \eqref{AOP_OjctFunctl} becomes 
\begin{multline}\label{AOP_OjctFunctlexpl}
\mathcal{P}[\g_w(t),\g_q(t),\l]= \frac{1}{\T}\tint
                  \sum_{\alpha,\beta =w,q} u_{\alpha\beta}\Biggl\{
                  \dot{\g}_\alpha(t) \g_\beta(t)\\
                 -\tauint \dot{\g}_\alpha(t)\dot{\g}_\beta(t-\tau)
                  e^{-2\mu\kappa_0\tau}\Biggr\}.
\end{multline}
Here, we introduced the coefficients 
\begin{equation}
\left(\!\begin{array}{cc}
u_{ww} & u_{wq}\\
u_{qw} & u_{qq}
\end{array}\!\right)\equiv
-2k_B T_{{\rm c}}^2
\left(\!\begin{array}{cc}
(\l-1)\chi^2\xi_w^2 & (\l-1)\chi\xi_w\xi_q\\
\l\bt\chi\xi_w\xi_q & \l\bt\xi_q^2
\end{array}\!\right)
\end{equation}
for notational simplicity.
The convolution type structure of \eqref{AOP_OjctFunctlexpl} naturally
suggests to solve the variational problem by a Fourier transformation.
We expand 
\begin{equation}\label{AOP_FourExps}
\g_\alpha(t)\equiv \sum_{n\in\mathbb{Z}} c^\alpha_n e^{in\omega t}
\quad\text{with}\quad
\omega\equiv 2\pi/\T
\end{equation}
and thus obtain 
\begin{multline}\label{AOP_OjctFunctlFour}
\mathcal{P}[\g_w(t),\g_q(t),\l]\\
=(-2\mu\kappa_0)
\sum_{\alpha,\beta=w,q}\sum_{n\in\mathbb{Z}} u_{\alpha\beta}
c_n^\alpha c_{-n}^{\beta} \frac{in\omega}{in\omega-2\mu\k_0}.
\end{multline}
Since \eqref{AOP_OjctFunctlFour} is quadratic in the Fourier
coefficients $c_n^\alpha$, it is straightforward to carry out the 
optimization with respect to $c_n^w$. 
Taking into account that the protocols must be real and therefore 
$c_{-n}^\alpha= c_n^{\alpha\ast}$, the conditions 
\begin{align}
& \partial_{c_n^w}\mathcal{P}[\g_w(t)\g_q(t),\lambda]\overset{!}{=}0,\\
& \partial_{c_n^{w\ast}}\mathcal{P}[\g_w(t),\g_q(t),\lambda]
  \overset{!}{=}0
\end{align}
yield 
\begin{equation}\label{AOP_FCoeff}
c_n^w= 
-\left(\frac{u_{wq}+u_{qw}}{2u_{ww}} + 2i\mu\k_0
\frac{u_{wq}-u_{qw}}{2n\omega u_{ww}}\right)c_n^q
\end{equation}
for $n\neq 0$.
Note that \eqref{AOP_OjctFunctlFour} does not depend on $c_0^\alpha$
and thus the optimal protocol will be unique only up to a trivial
offset $c_0^w$. 
To comply with the constraint \eqref{AOP_EffConstr}, the Lagrange 
multiplier $\lambda$ must be chosen such that
\begin{equation}\label{AOP_LMCond}
\partial_\l\mathcal{P}[\g_w(t),\g_q(t),\l]=0,
\end{equation}
where the derivative has to be taken before the $c_n^w$ are replaced
by the solution \eqref{AOP_FCoeff}.
After some algebra, \eqref{AOP_LMCond} reduces to the simple condition
\begin{equation}
\bt^2 - (\l-1)^2(\bt-1)^2=0,
\end{equation}
which is fulfilled for
\begin{equation}\label{AOP_LagrangeSol}
\l_\pm = (1-\bt\pm\bt)/(1-\bt).
\end{equation}
Inserting \eqref{AOP_LagrangeSol} and \eqref{AOP_FCoeff} into 
\eqref{AOP_OjctFunctlFour} yields
\begin{align}
\bar{P}_+ & \equiv\mathcal{P}[\g_w(t),\g_q(t),\l_+] =0,\\
\bar{P}_- & \equiv\mathcal{P}[\g_w(t),\g_q(t),\l_-]
            =8k_BT_{{\rm c}}^2\mu\k_0\xi_q^2\bt(1-\bt)
             \sum_{n=1}^\infty |c_n^q|^2\nonumber\\
& = \frac{k_B T_{{\rm c}}^2\mu\k_0}{\T}\bt(1-\bt)\tint
    \left(\g_q(t)- \bar{\g}_q\right)^2,
\end{align}
where $\bar{\g}_q$ is defined in \eqref{Ex_AbbrevDef}.
Consequently, the relevant solution for the Lagrange multiplier is
given by $\l_-$.
Finally, the optimal protocol $\g^\ast(t,\eta)$ is obtained by summing
up the Fourier series \eqref{AOP_FourExps}.
The explicit result \eqref{Ex_MaxPowProt} can be found by, first, 
evaluating
\begin{align}
& \dot{\g}_w^\ast(t,\eta)
  = i\omega\sum_{n\in\mathbb{Z}} n c_n^w e^{in\omega t}\nonumber\\
& = -\sum_{\substack{n\in\mathbb{Z}\\n\neq 0}}
    \left(\frac{in\omega(u_{wq}+u_{qw})}{2u_{ww}}
    -2\mu\k_0 \frac{u_{wq}-u_{qw}}{2 u_{ww}}\right) 
    c_n^q e^{in\omega t}\nonumber\\
& = -\frac{u_{wq}+u_{qw}}{2u_{ww}}\dot{\g}_q(t)
    +2\mu\k_0 \frac{u_{wq}-u_{qw}}{2 u_{ww}}
    \left(\g_q(t)-\bar{\g}_q\right)\nonumber\\
&=  -\frac{2k_BT_{{\rm c}}}{\chi}
    \Bigl(2\mu\k_0\bt\left(\g_q(t)-\bar{\g}_q\right)
    -(1-\bt)\dot{\g}_q(t)\Bigr)
\label{AOP_DiffEqOptProt}
\end{align}
and, second, solving the simple differential equation 
\eqref{AOP_DiffEqOptProt}.

\begin{acknowledgments}
K. B. acknowledges a short-term scholarship for PhD-students from the
German Academic Exchange Service (DAAD). 
K. S. was supported by MEXT (23740289).
\end{acknowledgments}

\end{document}